\documentclass[a4paper,12pt]{article}

\usepackage{amssymb,amsfonts,amsmath,epsfig,float,graphicx,xcolor}
\usepackage{mathrsfs}

\usepackage{parskip}

\usepackage{tgtermes}

\usepackage[superscript,biblabel]{cite}
\newcommand{\inlinecite}[1]{{\renewcommand\citemid{}\cite[\hspace*{0.8pt}]{#1}}}

\def\be{\begin{eqnarray}}
\def\ee{\end{eqnarray}}
\def\nnee{\nonumber\ee}

\newcommand{\upd}{{\rm d}}

\def\Co{{\mathbb C}}
\def\Mo{{\mathbb M}}

\def\Ro{{\mathbb R}}
\def\Io{{\mathbb I}}

\def\tinyapprox{{\mbox{\tiny approx}}}

\def\Im{\mbox{ Im }}
\def\Tr{\mbox{ Tr }}

\def\Aalg{\mathfrak{A}}
\def\Malg{\mathfrak{M}}
\def\henerg{\mathfrak{h}}
\def\kenerg{\mathfrak{k}}

\def\subOmega{{_\Omega}}
\def\subPhi{{_\Phi}}
\def\subPsi{{_\Psi}}

\def\POm{\mathscr P_\subOmega}
\def\DeltaOm{\Delta_\subOmega}
\def\SOm{S_\subOmega}

\newtheorem{definition}{Definition}
\newtheorem{proposition}{Proposition}
\newtheorem{theorem}{Theorem}
\newtheorem{lemma}{Lemma}

\newtheorem{corollary}{Corollary}

\def\beginproof{\par\strut\vskip 0.0cm\noindent{\bf Proof}\par}
\def\endproof{\par\strut\hfill$\square$\par\vskip 0.1cm}

\def\pbsize{11.5cm}

 \title{Exponential arcs in manifolds of quantum states}
\author{Jan Naudts\\
Universiteit Antwerpen\\
\small Physics Department, Universiteitsplein 1, 2610 Antwerpen, Belgium\\
}
\date{}

\begin{document}
\maketitle

\begin{abstract}

The manifold under consideration consists of the
faithful normal states on a sigma-finite von Neumann algebra in standard form.
Tangent planes and approximate tangent planes are discussed.
A relative entropy/divergence function
is assumed to be given.
It is used to generalize the notion of an exponential arc connecting one state to another.
The generator of the exponential arc is shown to be unique up to an additive constant. 
In the case of Araki's relative entropy every selfadjoint element
of the von Neumann algebra generates an exponential arc. The generators
of composed exponential arcs are shown to add up.
The metric derived from Araki's relative entropy is shown to reproduce
the Kubo-Mori metric. The latter is the metric used in Linear Response Theory.
The e- and m-connections describe a dual pair of geometries.
Any  finite number of linearly independent generators determines a submanifold of states
connected to a given reference state.
Such a submanifold is a quantum generalization of a dually flat statistical manifold.

 
\end{abstract}

\section{Introduction}

Goal of the present paper is to show that the theory of quantum statistical manifolds can be
formulated without reference to density matrices. It is tradition to describe the statistical
state of a quantum model by a density matrix. In many cases this suffices, in particular when the
Hilbert space of wavefunctions is finite-dimensional. However, even simple models
such as the quantum harmonic oscillator or the hydrogen atom
require an infinite-dimensional Hilbert space.  This involves handling of unbounded operators
which cause considerable technical complications. These complications are avoided in the present work.

A one-to-one correspondence between density matrices and quantum states is usually accepted.
The quantum states form the sample space of the statistical description.
An alternative description emerged in the past century. 
It introduces the notion of a mathematical state on an algebra of observables
which can be realised as an algebra of bounded operators on a Hilbert space.
See for instance \inlinecite{DJ64,DJ69,RD69,EG72,BR79}. 

Equilibrium states of Quantum Statistical Mechanics are described by the quantum analogue of
the probability distribution of Gibbs, which is a density matrix $\rho$ of the form
\be
\rho&=&\frac{1}{Z}e^{-\beta H}
\nnee
with $H$ a Hermitian matrix, $\beta$ a parameter called the inverse temperature
and $Z$ a function of $\beta$ used to normalize the density matrix $\rho$ so that its trace equals 1.
Models described in this way can be said to belong to a quantum exponential family.
They posses an intriguing property called the Kubo-Martin-Schwinger (KMS) condition
\cite{HHW67}.
The KMS condition describes a symmetry property of the time evolution of quantum states.
This symmetry coincides with the symmetry between left and right multiplication of operators
which is studied in Tomita-Takesaki theory\cite{TM70}. 
The book of Bratteli-Robinson\cite{BR79} can be used as a reference text for this theory.

The notion of a statistical manifold is studied in Information Geometry\cite{CNN72,EB75,AS85,AN00,AJLS18}.
It is a manifold of probability distributions. The quantum analogue is described in Chapter 7
of \inlinecite{AN00} as a manifold of $k$ by $k$ density matrices.
The book of Petz\cite{PD08} reviews several aspects of quantum statistics, including the basics of Quantum Information 
and Quantum Information Geometry.

The relative entropy of Umegaki\cite{UH62} is the starting point to implement Amari's dually flat geometry
on the quantum manifold. 
Note that a relative entropy is called a divergence function in the mathematical literature.
Araki\cite{AH73,AH74,AH76} 
generalized Umegaki's relative entropy to the context of mathematical states on an algebra
of bounded operators on a Hilbert space. The use of Araki's relative entropy replacing
that of Umegaki is the core of the present work.

Exponential arcs were introduced in \inlinecite{PC07,PG13} and used in \inlinecite{SST17}.
These arcs can be considered as one-parameter exponential families embedded in the manifold.
The maximal exponential model centered at a given probability distribution $p$
equals the set of all probability distributions connected to $p$ by an open exponential arc. 
Exponential arcs  were studied in the quantum setting by the author\cite{NJ22}.
Here the definition is generalized. The exponential arcs are used to define quantum statistical
manifolds as submanifolds of the manifold of all quantum states.

The Radon-Nikodym theorem plays an important role in probability theory.
For each measure absolutely continuous w.r.t.~the  reference measure there
exists an essentially unique probability distribution function.
The problem that arises in the non-commutative context is the non-uniqueness of the definition of
the Radon-Nikodym derivative. This leads to different definitions of the relative entropy
and of the exponential arcs.
First attempts to reformulate the theory of the quantum statistical manifold in terms of states
on a $C^*$-algebra are found in \inlinecite{NJ18,NJ18c} and in \inlinecite{NJ22}. 
These two approaches differ in the choice of Radon-Nikodym derivative.
In the present work the definition of an exponential arc
is generalized so that it depends explicitly on the choice of relative entropy
and in that way on the choice of Radon-Nikodym derivative. 

The alternative approach of \inlinecite{CIJM19} relies on Lie theory for the
group of bound\-ed operators with bounded inverse.
The state space is partitioned into the disjoint union of the orbits
of an action of the Lie group. Under mild conditions it is shown
that the orbits are Banach manifolds. The restriction to bounded operators
implies that the orbits do not connect quasi-equivalent states when
the Radon-Nikodym derivatives are unbounded operators.

\section{KMS states}
\label{sect:kms}

\hskip 2cm\parbox{\pbsize}
{\em
Equilibrium states of Quantum Statistical Mechanics satisfy the KMS condition.
In the GNS representation an equilibrium state becomes a faithful state on a $\sigma$-finite von Neumann algebra
of operators on a complex Hilbert space.
The state is defined by a normalized cyclic and separating vector in the Hilbert space.
}

The state of a model of Statistical Physics can be described by a mathematical state on a $C^*$-algebra $\Aalg$.
It can be represented by a normalized vector $\Omega$ (a {\em wave function})
in a Hilbert space $\mathscr H$. This is known as  the GNS (Gelfand Naimark Segal) representation theorem.
Observable quantities are represented by self-adjoint operators on $\mathscr H$. The quantum expectation
$\langle x\rangle$ of the operator $x$ is then given by
\be
\langle x\rangle &=&(x\Omega,\Omega),
\label{kms:vectorstate}
\ee
with in the r.h.s.~the scalar product of the two vectors $x\Omega$ and $\Omega$.
Note that the mathematics convention is followed that the scalar product (inner product)
is linear in its first argument and
conjugate-linear the second argument. In Dirac's {\em bra-ket} notation it reads
\be
\langle x\rangle &=&\langle\Omega|x\Omega\rangle.
\nnee

For convenience one works with a von Neumann algebra $\Malg$ of bounded operators on 
the Hilbert space $\mathscr H$. Observables of interest, when unbounded, are represented
by operators affiliated with $\Malg$. 
The state $\omega$ on the $C^*$-algebra extends to a vector state on $\Malg$
again denoted $\omega$. It is given by
\be
\omega(x)&=&(x\Omega,\Omega),
\qquad x\in \Malg.
\nnee
The vector $\Omega$ is cyclic for $\Malg$, which means that the subspace $\Malg\Omega$
is dense in the Hilbert space $\mathscr H$. It is also assumed in what follows that the state $\omega$
is faithful, i.e. $\omega(x^*x)=0$ implies $x=0$. This implies that $\Omega$ is a separating vector for
$\Malg$, i.e.~$x\Omega=0$ implies $x=0$ for any $x$ in $\Malg$,
and hence it is a cyclic vector for the commutant $\Malg'$ of $\Malg$, the algebra of all operators commuting
with all of $\Malg$.

Equilibrium states of Statistical Mechanics are characterised by the KMS (Kubo Martin Schwinger) condition
\cite{HHW67}.
Roughly spoken, this condition states that the quantum time evolution of the model
has an analytic extension into the complex plane. This is made more precise in what follows.

The time evolution is described by a strongly continuous one-parameter group $t\in\Ro\mapsto u_t$
of unitary operators which leave the algebra $\Malg$ unchanged, i.e.~$x\in \Malg$
implies that $x_t=u_t^*x u_t$ belongs to $\Malg$ for all $t$.
The operators $u_t$ are determined by a self-adjoint operator $H$
\be
u_t&=&e^{-itH}
\nnee
which is the generator of the time evolution in the GNS representation.
The time derivative of $x_t$ satisfies
\be
i\frac{\upd\,}{\upd t}x_t&=&[x_t,H]_-.
\nnee
This equation has the same form as Heisenberg's equation of motion.

The KMS condition requires that 
for any pair $x,y$ of operators in $\Malg$ there exists a complex function $F(w)$,
defined and continuous on the strip $-\beta\le \Im w\le 0$ and analytic inside with boundary values
\be
F(t)=(x_ty\Omega,\Omega)
\quad\mbox{ and }\quad 
F(t-i\beta)=(yx_t\Omega,\Omega),
\qquad t\in\Ro.
\nnee
In the Mathematics Literature the parameter $\beta$,
which is the inverse temperature of the model,
is usually taken equal to 1 or -1.

An immediate consequence of the KMS condition being satisfied
is that the state $\omega$ is invariant. Indeed, take $y$ equal to the identity operator.
Then one has $F(t-i\beta)=F(t)$ for all $t$ in $\Ro$. If in addition $x$ is self-adjoint then $F(t)$ is a real function.
From Schwarz reflection principle one then concludes that $F(w)$ is a constant function. This implies
$\omega(x_t)=\omega(x)$ for all self-adjoint $x$ and hence for all $x$. The GNS theorem then guarantees that
the vector $\Omega$ can be taken to be invariant, i.e.~$u_t\Omega=\Omega$ for all $t$.

\section{The modular operator}
\label{sect:modop}

\hskip 2cm\parbox{\pbsize}
{\em
The quantum-mechanical time evolution coincides with
the modular automorphism group of Tomita-Takesaki theory.
}

The KMS condition when satisfied expresses a symmetry which is present
in the context of non-commuting operators. The symmetry is the inversion of the order 
of multiplication of operators. In non-commutative groups the modular function links left and right
Haar measures. The analogue in Functional Analysis is studied in the theory of the modular operator,
also called Tomita-Takesaki theory \cite{TM70}.

The operator $e^{-\beta H}$, with $H$ the generator of the quantum time evolution
is traditionally denoted $\DeltaOm$.
It is the modular operator of Tomita-Takesaki theory.
It is in general an unbounded operator such that
$\Malg\Omega$ is in the domain of definition of the square root $\DeltaOm^{1/2}$ of $\DeltaOm$.
Hence, the expression
\be
F(w)&=&(x\DeltaOm^{i w/\beta}y\Omega,\Omega),
\qquad
x,y\in\Malg,
\ee
is well-defined for $0\ge\Im w\ge -\beta/2$.
The other half of the strip $0\ge \Im w\ge -\beta$ is covered by the Schwarz reflection principle.
Indeed, if $x$ and $y$ are selfadjoint then one can show with Tomita-Takesaki theory that the map
$t\mapsto F(t-i\beta/2)$ is a real function.
Hence, the principle can be applied to obtain $F(w)=\overline{F(w-i\beta)}$.

The unitary time evolution operators $u_t$ can be written as 
\be
u_t&=&\DeltaOm^{it/\beta}.
\nnee
The time evolution of an operator $x$ in the Heisenberg picture is then given by 
\be
x_t=\tau_t^{\scriptscriptstyle\Omega} x&=&\DeltaOm^{-it/\beta}\,x\,\DeltaOm^{it/\beta}.
\nnee
The action $t\mapsto \tau_t^{\scriptscriptstyle\Omega}$ of the group $\Ro,+$ is called the modular automorphism group.

The modular conjugation operator $J$ of Tomita-Takesaki theory represents the symmetry
which is at the basis of the theory. It is a conjugate-linear operator satisfying
$J=J^*$ and $J^2=\Io$.
An operator $x$ belongs to the von Neumann algebra $\Malg$ if and only if
$JxJ$ belongs to the commutant algebra $\Malg'$. The latter is the space of operators commuting with all
operators in $\Malg$.
The product $J\DeltaOm^{1/2}$ is denoted $\SOm$ and has the property that
\be
\SOm x\Omega=x^*\Omega,
\qquad
x\in\Malg.
\nnee

\section{Dual cones}

\hskip 2cm\parbox{\pbsize}
{\em
The natural positive cone $\POm$ is needed in subsequent sections.
One reason for making use of it is that there exists a
one-to-one correspondence between normal states on $\Malg$ and normalized vectors in $\POm$. 

}

Section 4 of \inlinecite{AH74} introduces the cones $V^\alpha_{_\Omega}$, $0\le \alpha\le 1/2$,  of vectors in $\mathscr H$.
The self-dual cone $V^{1/4}_{_\Omega}$ is called the {\em natural positive cone}
and is denoted $\POm$.

By definition is $V^\alpha_{_\Omega}$ the closure of the cone
\be
\{\DeltaOm^\alpha x\Omega:\, x\in\Malg, x\ge 0\}.
\nnee

The cone $V^{1/2}_{_\Omega}$ is used in \inlinecite{NJ22}
to introduce exponential arcs.
It is equal to the closure of the set
\be
\{y\Omega:\,y\in\Malg', y\ge 0\}.
\nnee
To see this note that
\be
\DeltaOm^{1/2}x\Omega&=&J\SOm x\Omega\cr
&=&
Jx^*\Omega\cr
&=&
y\Omega
\nnee
with $y=Jx^*J$. The latter is an arbitrary element of the commutant $\Malg'$.

The following characterisation of the natural positive cone $\POm$ is found in Section 2.5 of \inlinecite{BR79}.

\begin{proposition}
The cone $\POm=V^{1/4}_{_\Omega}$ equals the closure of the set of vectors
\be
\{xJx\Omega:\,x\in{\Malg}\}.
\label{man:conedef}
\ee
\end{proposition}

This result can be understood as follows.
Take $\Phi$ in $\mathscr P$ of the form (\ref {man:conedef}), i.e.~$\Phi=xJx\Omega$
with $x$ in $\Malg$.
Let
\be
y&=&\DeltaOm^{-1/4}x\DeltaOm^{1/4}.
\label{modop:temp}
\ee
This expression can be inverted to
\be
x&=&\DeltaOm^{1/4}y\DeltaOm^{-1/4}
\nnee
so that
\be
\Phi =xJx\Omega
&=&
\DeltaOm^{1/4}y\DeltaOm^{-1/4}\,J\,\DeltaOm^{1/4}y\DeltaOm^{-1/4}\Omega\cr
&=&
\DeltaOm^{1/4}yJ\,\DeltaOm^{1/2}y\Omega\cr
&=&
\DeltaOm^{1/4}y\SOm y\Omega\cr
&=&
\DeltaOm^{1/4}yy^*\Omega.
\nnee
Assume now that one could prove that the operator $y$
defined by (\ref {modop:temp}) belongs to $\Malg$
then the above calculation would show that $\Phi$ is of the form $\Phi=\DeltaOm^{1/4}a\Omega$
with $a=yy^*$ a positive element of $\Malg$.
The actual proof of the proposition uses that $\tau^\Omega_tx=\DeltaOm^{-it} x\DeltaOm^{it}$
belongs to $\Malg$.

The cone $\POm$ is independent \cite{AH74} of the choice of the cyclic and separating vector 
$\Omega$ in ${\POm}$ and the isometry $J$ is the same for all these choices.
For this reason it is said to be universal.

From (\ref {man:conedef}) it is easy to see that each vector in $\POm$ is an eigenvector 
with eigenvalue 1 of 
the modular conjugation operator $J$.  Indeed, one has
\be
xJx\Omega=x(JxJ)\Omega=(JxJ)x\Omega=J(xJx\Omega).
\nnee
Here, use is made of $J\Omega=\Omega$ and the fact that the operators $x$ and $JxJ$ commute with each other.

\section{A manifold of quantum states}

\hskip 2cm\parbox{\pbsize}
{\em
A manifold $\Mo$ of vector states on the von Neumann algebra $\mathcal{M}$ is defined.
Tangent vector fields are Fr\'echet-derivatives of paths in $\Mo$.

}

Introduce the notation $\omega_\subPhi$ for the vector state defined by the normalized vector $\Phi$
in $\mathscr H$. It is given by
\be
\omega_\subPhi(x)&=&(x\Phi,\Phi),
\qquad
x\in\Malg.
\nnee

A manifold $\Mo$ of states on the von Neumann algebra $\Malg$ is defined by
\be
\Mo&=&\{\omega_\Phi:\,\Phi\in{\mathscr P}_{_\Omega},\mbox{ normalized, cyclic and separating for }\Malg\}.
\nnee
The equilibrium state $\omega=\omega_\subOmega$ is taken as a reference point in $\Mo$. 
The subset ${\mathscr P}_{_\Omega}$ of $\mathscr H$ is the natural positive cone introduced in
the previous section.
 
The topology on the manifold $\Mo$ is that of the operator norm. One has
\be
||\omega_\subPhi-\omega_\subPsi||
&=&\sup\{|\omega_\subPhi(x)-\omega_\subPsi(x)|:\,x\in\Malg, ||x||\le 1\}.
\nnee
Several topologies can be defined on the algebra $\Malg$. Particularly relevant is the $\sigma$-weak topology.
For what follows it is important to know that in the present context
a state $\omega$ on $\Malg$ is said to be normal if and only if it is $\sigma$-weakly continuous
and if and only if it is a vector state. See for instance Theorems 2.4.21 and  2.5.31 of \inlinecite{BR79}.

Any tangent vector is a $\sigma$-weakly continuous linear functional on the von Neumann algebra $\Malg$.
Let $t\mapsto\gamma_t$ be a Fr\'echet-differentiable map defined on an open interval of $\Ro$
with values in the manifold $\Mo$.
The derivative
\be
\dot\gamma_t=\frac{\upd\,}{\upd t}\gamma_t
\nnee
is required to exist as a Fr\'echet derivative, i.e.~it satisfies
\be
||\gamma_{t}-\gamma_{s}-(t-s)\dot\gamma_t||&=&\mbox{o}(t-s).
\nnee
From the normalization $\gamma_t(1)=1$ for all $t$ in the domain of the map one obtains $\dot\gamma_t(1)=0$.
From $\gamma_t(x^*)=\overline{\gamma_t(x)}$ one obtains
$\dot\gamma_t(x^*)=\overline{\dot\gamma_t(x)}$. Hence, the linear functional $\dot\gamma_t$ is Hermitian.

There are several ways to define the tangent space $T_\omega\Mo$ at the point $\omega$ in $\Mo$.
The easiest way is to consider the manifold $\Mo$ as embedded in the Banach space $\Malg_*$ of 
$\sigma$-continuous Hermitian linear functionals on $\Malg$ (see Proposition 2.4.18 of \inlinecite{BR79}).
Then any Hermitian $\sigma$-weakly continuous linear function $\chi$ satisfying $\chi(1)=0$
defines a tangent curve $t\mapsto\omega+t\chi$ with Fr\'echet derivative $\chi$.
This justifies the definition that $T_\omega\Mo$ consists of all $\sigma$-weakly continuous 
Hermitian linear functions $\chi$ on $\Malg$ satisfying $\chi(1)=0$.

\section{Approximate tangents}
\label{sect:approx}

\hskip 2cm\parbox{\pbsize}
{\em Approximate tangent vectors can be defined in an intrinsic manner.
}

An alternative definition of the tangent space starts from the following observation.

\begin{proposition}
The set ${\mathscr T}_\omega$ defined by
\be
{\mathscr T}_\omega&=&
\{\lambda(\phi-\psi):\,\phi,\psi\in\Mo, \lambda\in\Ro\mbox{ and }\omega=\frac 12(\phi+\psi)\}.
\nnee
is a linear subspace of the tangent space $T_\omega\Mo$.
\end{proposition}

\beginproof
Let $\phi$ and $\psi$ be two states in $\Mo$ such that $\omega=\frac 12(\phi+\psi)$.
Construct a Fr\'echet-differentiable path $\gamma$ by
\be
\gamma_t&=&(1-t)\psi+t\phi,
\qquad
t\in(0,1).
\nnee
The state $\gamma_t$ belongs to the manifold $\Mo$ because the latter is a convex set.
In particular one has $\omega=\gamma_{1/2}$ and $\phi-\psi=\dot\gamma_{1/2}$ is a tangent vector.
This shows that $\phi-\psi$ and hence also $\lambda(\phi-\psi)$ belongs to $T_\omega\Mo$.
One concludes that ${\mathscr T}_\omega\subset T_\omega\Mo$.

Assume now that $\lambda(\phi-\psi)$ and $\lambda'(\phi'-\psi')$ both belong to ${\mathscr T}_\omega$.
We have to show that 
\be
\lambda(\phi-\psi)+\lambda'(\phi'-\psi') 
\nnee
belongs to ${\mathscr T}_\omega$.
If $\lambda=0$ or $\lambda'=0$ then the claim is clearly satisfied.
Without restriction assume $\lambda=1$.

If $\lambda'>0$ then choose
\be
\phi''=\frac 1{1+\lambda'}(\phi+\lambda'\phi')
\quad\mbox{ and }\quad
\psi''=\frac 1{1+\lambda'}(\psi+\lambda'\psi').
\nnee
Both $\phi''$ and $\psi''$ belong to $\Mo$ because the latter is a convex set.
One verifies that $\phi''+\psi''=2\omega$
and
\be
(1+\lambda')(\phi''-\psi'')
&=&
\phi-\psi+\lambda'(\phi'-\psi').
\nnee
This shows that the latter sum belongs to ${\mathscr T}_\omega$.

In the case that $\lambda'<0$ one chooses
\be
\phi''=\frac 1{1-\lambda'}(\phi-\lambda'\psi')
\quad\mbox{ and }\quad
\psi''=\frac 1{1-\lambda'}(\psi-\lambda'\phi')
\nnee
to reach the same conclusion.
This finishes the proof that ${\mathscr T}_\omega$ is a linear subspace of $T_\omega\Mo$.

\endproof

Introduce the notations
\be
{\mathscr R}_{\omega,\epsilon}&=&\bigcup{\mathscr T}_\phi
\mbox{ with }\phi\in\Mo
\mbox{ such that } ||\phi-\omega||<\epsilon \mbox.
\nnee
and
\be
{\mathscr T}_\omega^\tinyapprox&=&\bigcap_{\epsilon> 0}\overline{{\mathscr R}_{\omega,\epsilon}}.
\nnee

The construction of ${\mathscr T}_\omega^\tinyapprox$ is analoguous to the construction of
the {\em approximate tangent space} in Chapter 3 of \inlinecite{SL83}.
Clearly is ${\mathscr T}_\omega\subset {\mathscr T}_\omega^\tinyapprox$.
Further properties are derived below.

\begin{proposition}
If $\gamma$ is a Fr\'echet-differentiable path in $\Mo$
then $\dot\gamma_t$ belongs to ${\mathscr T}_\omega^\tinyapprox$ with $\omega=\gamma_t$.
\end{proposition}

\beginproof
Let $\gamma$ be a Fr\'echet-differentiable path in $\Mo$.
Without restriction of generality assume that $\gamma_0=\omega$.
For any $\epsilon>0$ and $\delta>0$ there exists $t\not=0$ such that
\be
||(\gamma_t+\gamma_{-t})/{2}-\gamma_0||&<&\epsilon.
\nnee
and
\be
||(\gamma_t-\gamma_{-t})/{2t}-\dot\gamma_0||&<&\delta.
\label{manif:temp}
\ee
Then $\phi$ defined by
\be
\phi&=&\frac 12(\gamma_t+\gamma_{-t})
\nnee
satisfies $||\phi-\omega||<\epsilon$
and $(\gamma_t-\gamma_{-t})/{2t}$ belongs to ${\mathscr R}_{\omega,\epsilon}$.
Hence,  (\ref {manif:temp}) shows that the tangent vector $\dot\gamma_0$ belongs to the closure 
of ${\mathscr R}_{\omega,\epsilon}$.
Because $\epsilon>0$ is arbitrary it belongs also to the intersection, which is ${\mathscr T}_\omega^\tinyapprox$.

\endproof

\begin{lemma}
${\mathscr R}_{\omega,\epsilon}$ is a linear subspace of $T_\omega\Mo$.
\end{lemma}

\beginproof
Take $\chi$ and $\xi$ in ${\mathscr R}_{\omega,\epsilon}$.
There exist $\phi$ and $\psi$ in $\Mo$ such that $\chi\in{\mathscr T}_\phi$
and $\xi\in{\mathscr T}_\psi$ with $||\phi-\omega||<\epsilon$ and $||\psi-\omega||<\epsilon$.
Therefore there exist real $\lambda,\mu$ and states $\phi_1,\Phi_2,\Psi_1,\Psi_2$ in $\Mo$
such that
\be
\chi=\lambda(\phi_1-\phi_2)
\quad\mbox{ and }\quad
\phi=\frac{1}{2}(\phi_1+\phi_2)
\nnee
and
\be
\xi=\mu(\psi_1-\psi_2)
\quad\mbox{ and }\quad
\psi=\frac{1}{2}(\psi_1+\psi_2).
\nnee
If $\lambda=0$ or $\mu=0$ then $\chi+\xi$ belongs to ${\mathscr R}_{\omega,\epsilon}$
without further argument. Assume therefore that $\lambda\not=0$ and $\mu\not=0$.
If $\lambda\mu>0$ then $\chi+\xi$ belongs to ${\mathscr T}_\pi$ with 
$\pi=(1-\alpha)\phi+\alpha\psi$ and $\alpha$ given by
\be
\alpha&=&\frac{\mu}{\lambda+\mu}.
\nnee
Indeed, let
\be
\pi_1&=&(1-\alpha)\phi_1+\alpha\psi_1,\cr
\pi_2&=&(1-\alpha)\phi_2+\alpha\psi_2.
\nnee
Then both $\pi_1$ and $\pi_2$ belong to $\Mo$ and satisfy 
\be
\pi_1+\pi_2&=&2(1-\alpha)\phi+2\alpha\psi\cr
&=&2\pi
\nnee
and
\be
(\lambda+\mu)(\pi_1-\pi_2)
&=&
(\lambda+\mu)\,[\frac{1-\alpha}{\lambda}\chi+\frac{\alpha}{\mu}\xi]\cr
&=&
\chi+\xi.
\nnee
In addition is
\be
||\pi-\omega||&=&||(1-\alpha)(\phi-\omega)+\alpha(\psi-\omega)||\cr
&\le&||(1-\alpha)(\phi-\omega)||+||\alpha(\psi-\omega)||\cr
&<&\epsilon.
\nnee
One concludes that in this case $\chi+\xi$ belongs to ${\mathscr R}_{\omega,\epsilon}$.

The case that $\lambda\mu<0$ is similar.
That $\chi\in{\mathscr R}_{\omega,\epsilon}$ implies $\lambda\chi\in{\mathscr R}_{\omega,\epsilon}$
is straightforward.
One concludes that ${\mathscr R}_{\omega,\epsilon}$ is a linear space.
It clearly is a subspace of $T_\omega\Mo$.

\endproof

\begin{proposition}
${\mathscr T}_\omega^\tinyapprox$ is a closed linear subspace of $T_\omega\Mo$.
\end{proposition}

\beginproof
The lemma shows that ${\mathscr R}_{\omega,\epsilon}$ is a linear subspace of $T_\omega\Mo$,
which is a space closed in norm. 
Hence, also the norm closure of ${\mathscr R}_{\omega,\epsilon}$ is a subset of this space and 
therefore also of $T_\omega\Mo$.

\endproof

\section{Majorized states}

\hskip 2cm\parbox{\pbsize}
{\em
The submanifold of states majorized by a multiple of the reference state $\omega$
is considered. 
}

\begin{definition}
A state $\phi$ on $\Malg$ is said to be majorized by a multiple of the state $\omega$
if there exists a positive constant $\lambda$ such that
\be
\phi(x^*x)&\le&\lambda\omega(x^*x)
\quad\mbox{ for all }
x\in\Malg.
\nnee
\end{definition}

Take $a'\not=0$ in the commutant algebra $\Malg'$ and let 
\be
\Phi=\frac 1{||a'\Omega||}a'\Omega.
\nnee
Then the state $\omega_\subPhi$ is majorized by a multiple of the state $\omega$.
Indeed, one has for any positive $x$ in $\Malg$
\be
\omega_\subPhi(x)
&=&\frac{(xa'\Omega,a'\Omega)}{(a'\Omega,a'\Omega)}\cr
&=&\frac{(a'^*a'x^{1/2}\Omega,x^{1/2}\Omega)}{(a'\Omega,a'\Omega)}\cr
&\le&\frac{||a'^*a'||}{(a'\Omega,a'\Omega)}\,\omega(x).
\nnee

It is well-known that all states majorized by a multiple of the state $\omega$
are obtained in this way. This is the content of the following proposition.

\begin{proposition}
\label{man:prop:commut}
If the vector state $\omega_\subPhi$ is majorized by a multiple of the state $\omega$
then there exists a unique element $a'$ of the commutant $\Malg'$ such that $\Phi=a'\Omega$.
\end{proposition}

\beginproof
An operator $a'$ is densely defined by
\be
a'x\Omega&=&x\Phi,
\qquad x\in\Malg.
\nnee
It satisfies $a'\Omega=\Phi$.
It is well-defined because $x\Omega=0$ implies
\be
||x\Phi||^2=\omega_\subPhi(x^*x)\le\mbox{constant }\omega(x^*x)=\mbox{constant }||x\Omega||^2=0
\nnee
so that $x\Phi=0$.

The operator $a'$ is bounded because
\be
||a'x\Omega||^2=\phi(x^*x)\le\mbox{constant }\omega(x^*x)=\mbox{constant }||x\Omega||^2.
\nnee

The operator $a'$ commutes with any $x$ in $\Malg$ because
\be
a'x(y\Omega)=xy\Phi=x(a'y\Omega)=xa'(y\Omega)
\nnee
and $\Omega$ is cyclic for $\Malg$.

The operator $a'$ is unique. Indeed, assume $b'$ in $\Malg'$ satisfies $\Phi=b'\Omega$.
Then one has for all $x$ in $\Malg$
\be
0=x(a'-b')\Omega=(a'-b')x\Omega.
\nnee
Hence, $a'-b'$ vanishes on $\Malg\Omega$ which is dense in the Hilbert space because $\Omega$
is cyclic for $\Malg$.
Because $a'-b'$ is a bounded and hence continuous operator it vanishes everywhere so that $a'=b'$.

\endproof

Item (8) of Theorem 3 of \inlinecite{AH74} implies the following.

\begin{proposition}
\label{rn:prop:exists}
If a vector state $\omega_\subPhi$, defined by a vector $\Phi$
in the natural positive cone $\POm$, is dominated by a multiple of the state $\omega$ then there
exists a unique element $a$ in the algebra $\Malg$ such that $\Phi=a\Omega$
and 
\be
\omega_\subPhi(x)=\omega(a^*xa),
\qquad
x\in\Malg.
\nnee
\end{proposition}

\beginproof
Proposition \ref{man:prop:commut} shows that $a'$ in the commutant $\Malg'$ exists such that
$\Phi=a'\Omega$. Because $\Phi$ and $\Omega$ both belong to $\POm$  one has
$\Phi=J\Phi=Ja'J\Omega$.

Let $a=Ja'J$. From $J\Malg'J=\Malg$ it follows that $a$ belongs to $\Malg$.
This shows the existence.

The element $a$ is unique because the correspondence between vector states on $\Malg$ and vectors in
$\POm$ is one-to-one and $\Omega$ is a separating vector for $\Malg$.

\endproof

If $\Malg$ is a commutative algebra then 
$a^*a$ is the Radon-Nikodym derivative of the state $\omega_\subPhi$
w.r.t.~the reference state $\omega$.

The submanifold of states of $\Mo$ majorized by a multiple of the state $\omega$
is dense in $\Mo$ in the sense that for any state $\phi$ in $\Mo$
there exists a sequence $(a_n)_n$ of elements of $\Malg$ with the property that $a_n\Omega$
is a Cauchy sequence and 
\be
\phi(x)&=&\lim_{n\rightarrow\infty}\omega(a_n^*xa_n),
\qquad
x\in\Malg.
\nnee
See Propositions 1.5 and 2.5 of \inlinecite{NG83}.

\begin{proposition}
A tangent vector $\chi$ 
belongs to the subspace ${\mathscr T}_\omega$ of the tangent space $T_\omega\Mo$
if and only if it is proportional to the difference of two states $\phi$ and $\psi$ in $\Mo$
both majorized by a multiple of the state $\omega$.
\end{proposition}

\beginproof

If $\chi$ belongs to ${\mathscr T}_\omega$  then by definition there exists states $\phi$ and $\psi$ in $\Mo$ 
such that  $\chi=\lambda(\phi-\psi)$ and $\phi+\psi=2\omega$.
The latter implies that both $\phi$ and $\psi$ are majorized by $2\omega$.

Conversely, assume that $\phi$ and $\psi$ in $\Mo$ both majorized by a multiple of the state $\omega$
and let $\chi=\lambda (\phi-\psi)$. This implies the existence of $\mu\ge 1$ and $\nu\ge 1$
such that $\phi\le\mu\omega$ and $\psi\le\nu\omega$.

Without restriction assume that $\lambda>0$.

Introduce
\be
\phi'=\omega+\rho\chi
\quad\mbox{ and }\quad
\psi'=\omega-\rho\chi
\nnee
with $\rho$ still to be chosen.
By construction is $\phi'+\psi'=2\omega$ and $\phi'-\psi'=2\rho\chi$.
Hence, if $\phi'$ and $\psi'$ are states in $\Mo$ and $\rho\not=0$ then
one can conclude that $\chi$ belongs to ${\mathscr T}_\omega$.

From
\be
\chi(x^*x)\le\lambda\phi(x^*x)\le\lambda\mu\omega(x^*x)
\nnee
and
\be
\chi(x^*x)\ge-\lambda\psi(x^*x)\ge-\lambda\nu\omega(x^*x)
\nnee
one obtains
\be
\phi'(x^*x)&\ge&[1-\rho\lambda\nu]\, \omega(x^*x),\cr
\psi'(x^*x)&\ge&[1-\rho\lambda\mu]\, \omega(x^*x).
\nnee
Choose now $\rho$ equal to the inverse of the maximum of $\lambda\mu$ and $\lambda\nu$
to prove the positivity of the functionals $\phi'$ and $\psi'$. Normalization
$\phi'(1)=\psi'(1)=1$ follows from $\chi(1)=0$. The functions are $\sigma$-weakly continuous as well.
Hence they are states in $\Mo$.
This ends the proof that $\chi$ belongs to ${\mathscr T}_\omega$.

\endproof

\section{Exponential arcs}
\label{sect:exparcs}

\hskip 2cm\parbox{\pbsize}
{\em
The author introduces\cite{NJ22} the notion of an exponential arc in Hilbert space,
inspired by the notion of exponential arcs in probability space as introduced by Pistone et al.\cite {PC07,PG13}.
Here, a definition is given which depends on the choice of a relative entropy.
}

In the present context a {\em divergence function} $D(\phi||\psi)$ is a real function of two states
$\phi$ and $\psi$ in the manifold $\Mo$.
It cannot be negative and it vanishes if and only if the two 
arguments are equal. A value of $+\infty$ is allowed.
An {\em energy function}
is an affine function $\henerg$
defined on a convex subset of the set of normal states on the algebra $\Malg$.

The following definition of an exponential arc in the manifold $\Mo$ assumes that a
divergence function $D(\phi||\psi)$ is given.

\begin{definition}
\label{arcs:def}
An exponential arc $\gamma$ is a path in the manifold 
\be
t\in[0,1]\mapsto\gamma_t\in\Mo
\nnee
for which there exists an energy function $\henerg$ such that
\begin{itemize}
 \item $\gamma_t$ is in the domain of $\henerg$;
 \item The divergence $D(\gamma_s||\gamma_t)$ between any two points of the arc is finite;
 \item For any state $\psi$ in the domain of $\henerg$ one has
 \be
D(\psi||\gamma_t)
&=&
D(\psi||\gamma_0)+D(\gamma_0||\gamma_t)
+t\,(\henerg(\gamma_0)-\henerg(\psi)),
 \qquad
 0\le t\le 1.\cr
 & &
 \label{arcs:arcdef}
 \ee
\end{itemize}

\end{definition}

The energy function $\henerg$ is the {\em generator} of the exponential arc.
The arc is said to connect the state $\gamma_1$ to the state $\gamma_0$.

A subclass of energy functions is formed by functions $\henerg$ for which there exists
a self-adjoint operator $h$ in the von Neumann algebra $\Malg$ so that
\be
\henerg(\psi)&=&\psi(h),
\qquad
\psi\in\Mo.
\label{arcs:specialh}
\ee
In such a case $h$ is called the generator as well.
The exponential arcs defined in \inlinecite{NJ22} agree with the above definition
with a generator defined by an unbounded operator affiliated with the
commutant algebra $\Malg'$.

\begin{proposition}
\label{arcs:prop:general}
Expression (\ref {arcs:arcdef}) implies
\be
 D(\gamma_s||\gamma_0)+D(\gamma_0||\gamma_s)&=&s\,(\henerg(\gamma_s)-\henerg(\gamma_0)).
 \label{arcs:propa}
 \ee
 and
 \be
D(\psi||\gamma_t)&=&D(\psi||\gamma_s)+D(\gamma_s||\gamma_t)+(t-s)\,(\henerg(\gamma_s)-\henerg(\psi)).
 \label{arcs:equiv}
 \ee 

\end{proposition}

Note that with $s=0$ expression (\ref {arcs:equiv}) reduces to  (\ref {arcs:arcdef}).

\beginproof

Take $\psi=\gamma_s$ in (\ref {arcs:arcdef}) to find
\be
D(\gamma_s||\gamma_t)
&=&
D(\gamma_s||\gamma_0)+D(\gamma_0||\gamma_t)
+t\,(\henerg(\gamma_0)-\henerg(\gamma_s)),
\qquad
0\le s, t\le 1.\cr
& &
\label{arcs:temp}
\ee
In particular, with $s=t$ this implies (\ref {arcs:propa}).

To prove (\ref {arcs:equiv}) use (\ref {arcs:temp}) to write the r.h.s.~as
\be
\mbox{r.h.s.}&=&
D(\psi||\gamma_s) +D(\gamma_s||\gamma_0)+D(\gamma_0||\gamma_t)\cr
& &\quad
+t\,(\henerg(\gamma_0)-\henerg(\psi))
-s\,(\henerg(\gamma_s)-\henerg(\psi)).
\nnee
Next, eliminate $D(\gamma_0||\gamma_t)$ 
and $D(\psi||\gamma_s)$
with the help of (\ref {arcs:arcdef}).
This gives
\be
\mbox{r.h.s.}
&=&
D(\psi||\gamma_s) +D(\gamma_s||\gamma_0)+D(\psi||\gamma_t)-D(\psi||\gamma_0)
-s\,(\henerg(\gamma_s)-\henerg(\psi))\cr
&=&
D(\gamma_s||\gamma_0)+D(\psi||\gamma_t)
+D(\gamma_0||\gamma_s)+s\,(\henerg(\gamma_0)-\henerg(\gamma_s))\cr
&=&
D(\psi||\gamma_t).
\nnee
To obtain the last line use (\ref {arcs:propa}).

\endproof

\begin{corollary}
\label{arcs:cor:subarcs}
If $t\mapsto\gamma_t$ is an exponential arc with generator $\henerg$ that connects $\gamma_1$ to $\gamma_0$
then for any $s,t$ in $[0,1]$ the map $\epsilon\mapsto\gamma_{(1-\epsilon)s+\epsilon t}$
is an exponential arc with generator $(t-s)\henerg$ that connects $\gamma_t$ to $\gamma_s$.
\end{corollary}

\begin{corollary}
If $t\mapsto\gamma_t$ is an exponential arc with generator $\henerg$ that connects $\gamma_1$ to $\gamma_0$
then $t\mapsto\gamma_{1-t}$ is an exponential arc with generator $-\henerg$, connecting the state $\gamma_0$
to the state $\gamma_1$.
\end{corollary}

The following two propositions deal with uniqueness of an exponential arc and of its generator.

\begin{proposition}
Let $\omega$ and $\phi$ be two states in $\Mo$. Fix  an energy function $\henerg$.
There is atmost one exponential arc $t\mapsto\gamma_t$ with generator $\henerg$ that
connects $\phi$ to $\omega$.
\end{proposition}

\beginproof
Assume both $t\mapsto\gamma_t$ and $t\mapsto\delta_t$ are 
exponential arcs connecting the state $\phi$ to the state $\omega$.
Subtract (\ref {arcs:arcdef}) from the same expression with $\gamma_t$ replaced by $\delta_t$
and take $s=0$.
This gives
\be
D(\psi||\delta_t)-D(\psi||\gamma_t)
&=&
D(\omega||\delta_t)-D(\omega||\gamma_t).
\label{arcs:prop:unique:temp}
\ee

Take $\psi$ equal to $\delta_t$. Then one obtains
\be
0\ge -D(\delta_t||\gamma_t)= 
D(\omega||\delta_t)-D(\omega||\gamma_t).
\nnee
On the other hand with $\psi=\gamma_t$ one obtains
\be
0\le 
D(\gamma_t||\delta_t)=
D(\omega||\delta_t)-D(\omega||\gamma_t).
\nnee
The two expressions together yield 
\be
D(\omega||\delta_t)-D(\omega||\gamma_t)
&=&0.
\nnee
This implies $D(\gamma_t||\delta_t)=0$.
By the basic property of a divergence one concludes that $\gamma_t=\delta_t$.

\endproof

\begin{proposition}
If the exponential arc $t\mapsto\gamma_t$ has two generators $\henerg$ and $\kenerg$ 
then these generators differ by a constant on their common domain of definition.
\end{proposition}

\beginproof
It follows from (\ref {arcs:arcdef}) that
\be
\henerg(\gamma_s)-\henerg(\psi)&=&\kenerg(\gamma_s)-\kenerg(\psi),
\qquad
s\in [0,1]
\label{arcs:prop2:temp}
\ee
for all states $\psi$ in the intersection of the domains of $\henerg$ and $\kenerg$.
This implies that a constant $c$ exists so that
\be
\kenerg(\psi)&=&\henerg(\psi)+c
\nnee
for all $\psi$ in the common domain.
\endproof

The requirement (\ref {arcs:arcdef}) is a stability condition. The generator $\henerg$ is a perturbation
which shifts the state $\gamma_0$ to the state $\psi$. 
This interpretation will become clear further on.
The effect on the relative entropy of the shift
along the arc $t\mapsto \gamma_t$ is linear. In the standard case
the relative entropy is based on the logarithmic function.
This justifies to call the path $t\mapsto\gamma_t$ an exponential arc. 

Note that the Pythagorean relation \cite{CI75,CM12}
\be
D(\psi||\gamma_t)&=&D(\psi||\gamma_s)+D(\gamma_s||\gamma_t)
\nnee
is satisfied for all $\psi$ with the same {\em energy} as the state $\gamma_s$, i.e.~with
\be
\henerg(\psi)=\henerg(\gamma_s).
\nnee
If the divergence function is interpreted as the square of a pseudo-distance then the above relation 
states that for an arbitrary state $\psi$
the point $\gamma_s$ of the arc which has the same energy is the point with minimal distance.

\section{The scalar potential}
\label{sect:scalar}

\hskip 2cm\parbox{\pbsize}
{\em
The exponential arc has a dual structure similar to that
found in Information Geometry \cite{AS85,AN00}.
}

Given an exponential arc $t\mapsto\gamma_t$ with generator $\henerg$
introduce the potential $\Phi_\gamma$ defined by
\be
\Phi_\gamma(t)&=&D(\gamma_0||\gamma_t)+t\henerg(\gamma_0).
\nnee
Its Legendre transform is given by
\be
\Phi^*_\gamma(\alpha)&=&\sup\{\alpha t-\Phi_\gamma(t):\,0\le t\le 1\}.
\nnee

\begin{proposition}
For any exponential arc $t\mapsto\gamma_t$ with generator $\henerg$ one has
\begin{itemize}
 \item [(a) ] The function $t\mapsto\henerg(\gamma_t)$ is strictly increasing;
 \item [(b) ] $\Phi_\gamma(t)=\Phi_\gamma(s)+D(\gamma_s||\gamma_t)+(t-s)\henerg(\gamma_s)$;
 \item [(c) ] The line $t\mapsto \Phi_\gamma(s)+(t-s)\henerg(\gamma_s)$ is tangent to the potential
 $\Phi_\gamma$ at the point $t=s$; this implies that the potential $\Phi_\gamma(s)$ is a strictly
 convex function, continuous on the open interval $(0,1)$;
  \item [(d) ] The following identity holds:
\be
\Phi_\gamma(s)+\Phi^*_\gamma(\henerg(\gamma_s))
=s\henerg(\gamma_s),
\qquad s\in[0,1].
\nnee
\end{itemize}

\end{proposition}

\beginproof

{\bf (a)}

Take $\psi=\gamma_t$ in (\ref {arcs:arcdef}). This gives
\be
0 &=&
D(\gamma_t||\gamma_s)+D(\gamma_s||\gamma_t)+(t-s)\left(\henerg(\gamma_s)-\henerg(\gamma_t)\right).
\nnee
Because divergences cannot be negative
this implies that $t\mapsto \henerg(\gamma_t)$ is non-dec\-reas\-ing.
Assume now that $\henerg(\gamma_s)=\henerg(\gamma_t)$.
Then it follows that
\be
0 =D(\gamma_t||\gamma_s)=D(\gamma_s||\gamma_t).
\nnee
The latter implies that $s=t$. One concludes that $s<t$ implies 
a strict inequality $\henerg(\gamma_s)<\henerg(\gamma_t)$.

{\bf (b)}

From the definition of the exponential arc one obtains
\be
D(\gamma_s||\gamma_t)+(t-s)\henerg(\gamma_s)
&=&
D(\psi||\gamma_t)-D(\psi||\gamma_s)+(t-s)\henerg(\psi).
\nnee
Take $\psi=\gamma_0$ in this expression to find
\be
D(\gamma_s||\gamma_t)+(t-s)\henerg(\gamma_s)
&=&
D(\gamma_0||\gamma_t)-D(\gamma_0||\gamma_s)+(t-s)\henerg(\gamma_0)\cr
&=&
\Phi_\gamma(t)-\Phi_\gamma(s).
\nnee

{\bf (c)}

From (b) one obtains
\be
\Phi_\gamma(t)-t\henerg(\gamma_s)&\ge&\Phi_\gamma(s)-s\henerg(\gamma_s),
\qquad
0\le t\le 1
\label{pot:temp}
\ee
because $D(\gamma_s||\gamma_t)\ge 0$ with equality if and only if $s=t$.
This implies that $t\mapsto \Phi_\gamma(s)+(t-s)\henerg(\gamma_s)$ is a line  tangent to the potential
$\Phi_\gamma(s)$.
By (a) the slope of this line is a strictly increasing function of $s$. Hence, the potential $\Phi_\gamma(s)$ 
is a strictly convex function, continuous on the open interval $(0,1)$.

{\bf (d)}

(\ref {pot:temp}) implies that
\be
\Phi^*_\gamma(\henerg(\gamma_s))
&=&
\sup_t \,t\henerg(\gamma_s)-\Phi_\gamma(t)\cr
&\le&
s\henerg(\gamma_s)-\Phi_\gamma(s).
\nnee
On the other hand one has using (b)
\be
\Phi_\gamma(s)+\Phi^*_\gamma(\henerg(\gamma_s))
&\ge&
\Phi_\gamma(s)+t\henerg(\gamma_s)-\Phi_\gamma(t)\cr
&=&
t\henerg(\gamma_s)-\left[D(\gamma_s||\gamma_t)+(t-s)\henerg(\gamma_s)\right]\cr
&=&
-D(\gamma_s||\gamma_t)+s\henerg(\gamma_s).
\nnee
The optimal choice $t=s$ yields the lower bound $s\henerg(\gamma_s)$.

\endproof

A dual parameter $\eta$ of the exponential arc $\gamma$, dual to the parameter $t$,
is the value $\henerg(\gamma_t)$ of the generator $\henerg$.
By item (a) of the proposition it is a strictly increasing function of $t$.
It is almost everywhere equal to the derivative $\dot\Phi_\gamma(t)$ of the value
of the potential along the path.

\section{The matrix case}

\hskip 2cm\parbox{\pbsize}
{\em
If $\rho$ and $\sigma$ are two density matrices then the obvious definition of an exponential
arc connecting $\sigma$ to $\rho$ is
\be
t &\mapsto&\sigma_t=\exp\left(\log\rho+t(\log\sigma-\log\rho)-\zeta(t)\right)
\nnee
with normalization $\zeta(t)$ given by
\be
\zeta(t)&=&\log\,\Tr \exp\left(\log\rho+t(\log\sigma-\log\rho)\right).
\nnee
It is shown below that the corresponding states given by 
\be
\phi_t(x)&=&\Tr\sigma_t x,
\qquad
x\in \Aalg
\nnee
form an exponential arc for the relative entropy of Umegaki \cite{UH62}
in the GNS-representation of the state $\omega_0$.
}

Fix a non-degenerate density matrix $\rho$ of size $n$-by-$n$. It is a positive-definite matrix
with trace $\Tr\rho$ equal to 1.

Umegaki's relative entropy for the pair of density matrices $\sigma$, $\tau$ is given by
\be
D(\sigma||\tau)&=&\Tr\sigma(\log\sigma-\log\tau).
\nnee

Assume now a map
\be
t\mapsto \sigma_t&=&\exp\left(\log\rho+th-\zeta(t)\right)
\label{matrix:exparc}
\ee
with normalization $\zeta(t)$ and $h$ given by 
\be
h&=&\log\sigma-\log\rho.
\nnee
This is the obvious definition of an exponential arc in terms of density matrices.
The corresponding potential is
\be
\Phi_\sigma(t)&=&D(\sigma_0||\sigma_t)+t\henerg(\sigma_0)\cr
&=&\zeta(t).
\nnee

The map (\ref {matrix:exparc}) is also an exponential arc in the sense of Definition \ref {arcs:def}.
To see this consider any density matrix $\tau$ and calculate
\be
D(\tau||\sigma_t)-D(\tau||\sigma_s)-D(\sigma_s||\sigma_t)
&=&
\Tr\tau(\log\tau-\log\sigma_t)\cr
& &-\Tr\tau(\log\tau-\log\sigma_s)\cr
& &-\Tr\sigma_s(\log\sigma_s-\log\sigma_t)\cr
&=&
-(t-s)\Tr(\tau-\sigma_s)h\cr
&=&
(t-s)\,(\henerg(\gamma_s)-\henerg(\tau))
\nnee
with 
\be
\henerg (\tau)=\Tr\tau h=\Tr\tau(\log\sigma -\log\rho).
\nnee
This is of the form (\ref {arcs:arcdef})
except that the relative entropy is
expressed in terms of density matrices in $\Malg$ instead of vector states in the
GNS representation of the state defined by the density matrix $\rho$.

An explicit construction of the GNS representation is possible.
See for instance the appendix of \inlinecite{NJ18}.
Let $\omega$ denote the state determined by the density matrix $\rho$
\be
\omega(A)&=&\Tr\rho A
\nnee
for any $n$-by-$n$ matrix $A$ with entries in $\Co$.
Such a matrix $A$ is represented on the Hilbert space $\mathscr{H}=\Co^n\otimes\Co^n$
by the operator $A\otimes\Io$, where $\Io$ is the $n$-by-$n$ identity matrix.
The von Neumann algebra $\Malg$ is the space of operators $A\otimes\Io$.

The matrix $\rho$ can be diagonalized. This gives the spectral representation
\be
\rho&=&\sum_i\,p_ie_i,
\nnee
where $(e_i)i_i$ is an orthonormal basis in $\Co^n$.
Let 
\be
\Omega&=&\sum_i\sqrt{p_i}e_i\otimes e_i.
\nnee
It is a normalized vector in $\mathscr H$. One readily verifies that
\be
\omega(A)&=&(A\otimes\Io\,\Omega,\Omega)
\nnee
for any $n$-by-$n$ matrix $A$.
In this way any density matrix $\rho$ defines a vector $\Omega$ in $\mathscr H$.
The vector $\Omega$ is cyclic and separating for $\Malg$ if $\rho$ is non-degenerate.
Hence, there is a one-to-one correspondence between non-degenerate density matrices and states in 
the manifold $\Mo$.
It is then straightforward to replace density matrices by states in the expressions obtained in
the first part of this section.

\section{The relative modular operator}

\hskip 2cm\parbox{\pbsize}
{\em
Araki\cite{AH77} introduces the relative modular operator $\Delta_{\Phi,\Psi}$
for any pair of vectors $\Phi$, $\Psi$ in the natural positive cone $\mathscr P$.
}

Assume that $\Phi$ and $\Psi$ are vectors in $\mathscr P$ which are separating for the algebra $\Malg$.
Then a conjugate-linear operator is defined by
\be
x\Psi\mapsto x^*\Phi,
\qquad
x\in\Malg.
\nnee
It is well-defined because by assumption $x\Psi=0$ implies that $x=0$ so that also $x^*\Phi=0$.
It is a closable operator. Indeed, assume the sequence $x_n\Psi$ converges to 0.
Then one has for any $y$ in the commutant $\Malg'$ that
\be
(x_n\Psi,y\Phi)=(y\Psi,x_n^*\Phi)
\nnee
converges to zero. By assumption is $\Psi$ separating for $\Malg$
so that it is cyclic for the commutant $\Malg'$. Hence, if the sequence $x_n^*\Phi$ converges then
it converges to 0. This shows closability of the operator.

Let $S_{\Phi,\Psi}$ denote the closure of this operator.
It satisfies
\be
S_{\Phi,\Psi}x\Psi&=&x^*\Phi,
\qquad
x\in\Malg.
\nnee
Its inverse equals $S_{\Psi,\Phi}$.

The relative modular operator $\Delta_{\Phi,\Psi}$ is defined by
\be
\Delta_{\Phi,\Psi} &=&S^*_{\Phi,\Psi}S_{\Phi,\Psi}.
\nnee

Important properties of the relative modular operator are
\be
\Delta_{\Phi,\Phi}=\Delta_\Phi
\quad\mbox{ and }\quad
S_{\Phi,\Psi}=J\Delta_{\Phi,\Psi}^{1/2}.
\nnee
where $J$ is the modular conjugation operator for the vector $\Phi$.

\section{Araki's relative entropy}
\label{sect:relent}

\hskip 2cm\parbox{\pbsize}
{\em
Araki\cite{AH76,AH77} uses the relative modular operator $\Delta_{\Phi,\Psi}$
to define the {\em relative entropy/divergence} $D(\phi||\psi)$
 of the corresponding states $\phi=\omega_\Phi$ and $\psi=\omega_\Psi$ by
\be
D(\phi||\psi)&=&((\log\Delta_{\Phi,\Psi})\Phi,\Phi).
\label{relent:relentdef}
\nnee
}

\begin{proposition}
The divergence $D(\phi||\psi)$ satisfies $D(\phi||\psi)\ge 0$ with equality
if and only if $\phi=\psi$.
\end{proposition}

\beginproof

Let 
\be
\Delta_{\Phi,\Psi}&=&\int\lambda\,\upd E_\lambda
\nnee
denote the spectral decomposition of the operator $\Delta_{\Phi,\Psi}$.
From the concavity of the logarithmic function it follows that
\be
D(\phi||\psi) 
&=&-((\log\Delta^{-1}_{\Phi,\Psi})\Phi,\Phi)\cr
&=&
-\int\log\lambda^{-1} \,\upd (E_\lambda\Phi,\Phi)\cr
&\ge&
-\log \int \lambda^{-1} \,\upd (E_\lambda\Phi,\Phi)\cr
&=&
-\log (\Delta^{-1/2}_{\Phi,\Psi}\Phi,\Delta^{-1/2}_{\Phi,\Psi}\Phi)\cr
&=&
-\log (\Psi,\Psi)\cr
&=&0.
\nnee
This shows that the divergence cannot be negative.

If $\phi=\psi$ then one has 
\be
D(\phi||\phi)=((\log\Delta_\Phi)\Phi,\Phi)=0
\nnee
because $\Delta_\phi\Phi=\Phi$.

Finally, $D(\phi||\psi)=0$ implies that $\Phi$ is in the domain of $\log\Delta_{\Phi,\Psi}$
and that 
$$\log\Delta_{\Phi,\Psi}\Phi=0.$$
The latter implies that 
\be
\Psi&=&\Delta^{-1}_{\Phi,\Psi}\Phi\,=\,\Phi.
\nnee
This shows that $D(\phi||\psi)=0$ vanishes only when $\Phi=\Psi$.

\endproof

Theorem 2.4 of \inlinecite{AH77} shows that
\be
\log\Delta_{\Phi,\Psi}+J\log\Delta_{\Psi,\Phi}J&=&0.
\nnee
Because $\Phi$ belongs by assumption to the natural positive cone $\mathscr  P$ it satisfies $\Phi=J\Phi$.
Hence one has also
\be
D(\phi||\psi)&=&-((\log\Delta_{\Psi,\Phi})\Phi,\Phi).
\nnee

\section{A theorem}
\label{sect:theorem}

\hskip 2cm\parbox{\pbsize}
{\em
Each self-adjoint element $h$ of the von Neumann algebra $\Malg$ defines an exponential arc
with as generator the energy function defined by $h$.

}

Araki\inlinecite{AH73} constructs for each selfadjoint operator $h$ in $\Malg$ 
a vector $\Phi_h$ in the natural positive cone $\mathscr P$
and calls $h$ the relative hamiltonian. 
Inspection of the explicit expression used in \inlinecite{AH73} shows that
\be
\Phi_h&=&\Omega+X h\Omega +\mbox{O}(h^2)
\label{relent:expans}
\ee
with the operator $X$ given by
\be
X
&=&
\int_0^{1/2}\upd u\,\Delta_{_\Omega}^{u}.
\nnee

The vector $\Phi_h$ defines a state $\phi_h$ by 
\be
\phi_h(x)&=&e^{-\xi(h)}(x\Phi_h,\Phi_h),
\qquad
x\in\Malg.
\nnee
Here, $\xi(h)$ is the normalization
\be
\xi(h)&=&\log (\Phi_h,\Phi_h).
\nnee

Theorem 3.10 of \inlinecite{AH77} implies that the state $\phi_h$ obtained in this way satisfies
for all $\psi$ in $\Mo$
\be
D(\psi||\phi_h)&=&D(\psi||\omega)-\psi(h)+\xi(h).
\label {theo:temp0}
\ee
Take $\psi=\phi_h$, respectively $\psi=\omega$, to find that the  normalization $\xi(h)$ is given by
\be
\xi(h)=\phi_h(h)-D(\phi_h||\omega)=\omega(h)+D(\omega||\phi_h).
\nnee

Consider now the path $\gamma$ defined by $\gamma_t=\phi_{th}$.
Then (\ref {theo:temp0}) becomes
\be
D(\psi||\gamma_t)&=&D(\psi||\omega)-t\psi(h)+\zeta(t).
\label{theo:temp}
\ee
with
\be
\zeta(t)\,=\,t\gamma_t(h)-D(\gamma_t||\omega)\,=\,t\omega(h)+D(\omega||\gamma_t)=\xi(th).
\nnee
From this last expression one obtains
\be
0\,\le\,
D(\gamma_t||\omega)+D(\omega||\gamma_t)&=&t[\gamma_t(h)-\omega(h)].
\nnee
From (\ref {relent:expans}) it follows that 
$\gamma_t$ converges to $\omega$ as $t\downarrow 0$. 
Hence, $D(\gamma_t||\omega)$ and $D(\omega||\gamma_t)$ converge to 0 faster than $t$.
This implies that the derivative $\dot\zeta(0)$
exists and equals $\omega(h)$.  This implies also that
\be
\frac{\upd\,}{\upd t}\bigg|_{t=0}D(\psi||\gamma_t)&=&\omega(h)-\psi(h).
\label{relent:deriv}
\ee

Elimination of $\zeta(t)$ from (\ref {theo:temp}) yields
\be
D(\psi||\gamma_t)&=&D(\psi||\omega)+D(\omega||\gamma_t)+t(\omega(h)-\psi(h)).
\nnee
This shows that $\gamma$ is an exponential arc connecting $\gamma_1$ to $\gamma_0=\omega$.

\begin{proposition}
One has
\be
\dot\gamma_0(x)
&=&
(T_{_\Omega} h\Omega,T_{_\Omega}  [x-\omega(x)]^*\Omega)
\label{relent:initial2}
\ee
with the operator $T_{_\Omega}$ given by
\be
T_{_\Omega} &=&\left(\frac{\Delta_{_\Omega}-1}{\log\Delta_{_\Omega}}\right)^{1/2}.
\label{relent:Tdef}
\ee
\end{proposition}

Note that this operator $T_{_\Omega}$ was introduced in \inlinecite{NVW75}.

\beginproof

From (\ref {relent:expans}) one obtains 
\be
\dot\gamma_0(x)=
\frac{\upd\,}{\upd t}\bigg|_{t=0}\gamma_t(x)
&=&
(xXh\Omega,\Omega)+(x\Omega,Xh\Omega)-\dot\zeta(0)\omega(x).
\label{relent:initial}
\ee
Write
\be
(xX h\Omega,\Omega)
&=&
\int_0^{1/2}\upd u\,(x\Delta_{_\Omega}^{u} h\Omega,\Omega)\cr
&=&
\int_0^{1/2}\upd u\,(\Delta_{_\Omega}^{u/2} h\Omega, \Delta_{_\Omega}^{u/2}x^*\Omega)
\nnee
and
\be
(x\Omega,Xh\Omega)
&=&\int_0^{1/2}\upd u\,(x\Omega,\Delta_{_\Omega}^u h\Omega)\cr
&=&\int_0^{1/2}\upd u\,(\Delta_{_\Omega}^{u/2}J\Delta_{_\Omega}^{1/2} x^*\Omega,
\Delta_{_\Omega}^{u/2}J\Delta_{_\Omega}^{1/2} h\Omega)\cr
&=&\int_0^{1/2}\upd u\,(J\Delta_{_\Omega}^{(1-u)/2}x^*\Omega,J\Delta_{_\Omega}^{(1-u)/2} h\Omega)\cr
&=&\int_{1/2}^1\upd u\,(J\Delta_{_\Omega}^{u/2}x^*\Omega,J\Delta_{_\Omega}^{u/2} h\Omega)\cr
&=&\int_{1/2}^1\upd u\,(\Delta_{_\Omega}^{u/2} h\Omega,\Delta_{_\Omega}^{u/2} x^*\Omega).
\nnee
The two contributions to (\ref {relent:initial}) can now be taken together. One obtains
\be
\dot\gamma_0(x)
&=&
\int_{0}^1\upd u\,(\Delta_{_\Omega}^{u/2} h\Omega,\Delta_{_\Omega}^{u/2} x^*\Omega)
-\dot\zeta(0)\omega(x)\cr
&=&
(T_{_\Omega} h\Omega,T_{_\Omega}  x^*\Omega)-\dot\zeta(0)\omega(x).
\nnee
Take $x=1$ to see that
\be
\dot\zeta(0)=(T_{_\Omega} h\Omega,T_{_\Omega} \Omega)=\omega(h)
\nnee
so that (\ref {relent:initial2}) follows.

\endproof

In summary, one has

\begin{theorem}
\label{relent:theorem}
Let $\omega$ in $\Mo$ be a vector state with cyclic and separating vector $\Omega$.
Choose the divergence function equal to the relative entropy
of Araki as defined by (\ref {relent:relentdef}).
For each self-adjoint element $h$ in $\Malg$ an energy function $\henerg$ is defined by
$\henerg(\phi)=\phi(h)$ and there exists an exponential arc $\gamma$
with generator $\henerg$ connecting some state $\gamma_1$ of $\Mo$ to the state $\gamma_0=\omega$. 
For any state $\psi$ in $\Mo$ the derivative of $D(\psi||\gamma_t)$ at $t=0$ exists and is given by
$\omega(h)-\psi(h)$. The derivative of the exponential arc at $t=0$ satisfies (\ref {relent:initial2}).
\end{theorem}

Further properties hold for the exponential arc of the above theorem.

\begin{proposition}
For any exponential arc $\gamma$ constructed in Theorem \ref {relent:theorem} the derivative
$\dot\gamma_0$ is a Fr\'echet derivative.
\end{proposition}

\beginproof
Let $\Xi(h)$ denote the remainder of order $h^2$ in (\ref {relent:expans}), i.e.
\be
\Phi_h&=&\Omega+X h\Omega +\Xi(h).
\nnee
Then one has using (\ref {relent:initial2})
\be
&&
\gamma_t(x)-\gamma_0(x)-t\dot\gamma_0(x)\cr
&=&
e^{-\zeta(t)}(x\Phi_{th},\Phi_{th})-\omega(x)-t(xX h\Omega,\Omega)-t(x\Omega,X h\Omega)
+t\omega(h)\omega(x)\cr
&=&
\left[e^{-\zeta(t)}-1+t\omega(h)\right]\omega(x)\cr
&+&
t\left[e^{-\zeta(t)}-1\right]\,
\left[(xX h\Omega,\Omega)+(x\Omega,X h\Omega)
\right]\cr
&+&
e^{-\zeta(t)}
\bigg[(x\Xi(th),\Omega)+(x\Omega,\Xi(th))+t^2(xXh\Omega,Xh\Omega)\cr
& &\qquad
+t(x\Xi(th),Xh\Omega)+t(xXh\Omega,\Xi(th))
+(x\Xi(th),\Xi(th))
\bigg].
\nnee
This yields
\be
||\gamma_t-\gamma_0-t\dot\gamma_0||
&\le&
|e^{-\zeta(t)}-1+t\omega(h)|\cr
& &+2t|e^{-\zeta(t)}-1|\,||Xh\Omega||\cr
& &+2e^{-\zeta(t)}\,||\Xi(th)||\cr
& &+e^{-\zeta(t)}\,\left[||\Xi(th)||+t||Xh\Omega||\right]^2.
\nnee
Each of the terms in the r.h.s.~of this expression is of order less than $t$ as $t$ tends to 0.
Hence, $\dot\gamma_0$ is a Fr\'echet derivative.

\endproof

\begin{proposition}[Additivity of generators]
\label{relent:prop:additive}
If the state $\phi$ is connected to the state $\omega$ by the exponential arc
with generator $h$ and $\psi$ is connected to $\phi$ by the exponential arc with generator $k$
then $\psi$ is connected to $\omega$ by the exponential arc with generator $h+k$
and $\omega$ is connected to $\psi$ by the exponential arc with generator $-h$.
\end{proposition}

For the proof see Proposition 4.5 of \inlinecite{AH73}.

\section{The metric}
\label{sect:metric}

\hskip 2cm\parbox{\pbsize}
{\em
Eguchi\cite{ES04} introduces the technique of deriving the metric of the tangent space
by taking two derivatives of the divergence.
Application here yields the metric which is used in the Kubo-Mori theory of linear response
\cite{KR57,MH65}.
}

Consider two exponential arcs $t\mapsto\gamma_t$ and $s\mapsto\eta_s$ with 
respective generators $h$ and $k$. They connect the states $\gamma_1$ and $\eta_1$ to the
reference state $\omega$.
The tangent vectors at $s=t=0$ are $\dot\gamma_0$ and $\dot\eta_0$. They belong to 
the tangent space $T_\omega\Mo$.
The scalar product between them is by definition given by
\be
(\dot\eta_0,\dot\gamma_0)_\omega
&=&-\frac{\partial\,}{\partial s}\frac{\partial\,}{\partial t}\bigg|_{s=t=0}
D(\eta_s||\gamma_t).
\nnee

Assume now that these exponential arcs are those constructed in Theorem \ref {relent:theorem}.
Then one has
\be
(\dot\eta_0,\dot\gamma_0)_\omega
&=&-\frac{\partial\,}{\partial s}\bigg|_{s=t=0}
\left(\omega(h)-\eta_s(h)\right)\cr
&=&
\dot\eta_0(h)\cr
&=&(T_{_\Omega} k\Omega,T_{_\Omega} (h-\omega(h))\Omega)\cr
&=&(T_{_\Omega} (k-\omega(k))\Omega,T_{_\Omega} (h-\omega(h))\Omega),
\label{metric:temp}
\ee
with the operator $T_{_\Omega} $ defined by (\ref {relent:Tdef}).
Note that in most applications one assumes that the expectations $\omega(h)$ of the generator $h$
and $\omega(k)$ of the generator $k$ vanish.
Then the result obtained here coincides with that used in \inlinecite{NVW75}.
In what follows a non-vanishing expectation of the generators is
taken into account.

Let us now discuss some technical issues.
The scalar product is well-defined by (\ref {metric:temp}). This follows from

\begin{lemma}
If two exponential arcs with initial point $\omega$ with generators $h$, respectively $k$, both in $\Malg$,
have the same initial tangent vector then one has
\be
T_{_\Omega} (h-\omega(h))\Omega&=&T_{_\Omega} (k-\omega(k))\Omega.
\nnee
\end{lemma}

\beginproof
Let $\gamma$ and $\eta$ be two exponential arcs with generators $h$, respectively $k$ in $\Malg$
and such that $\gamma_0=\eta_0=\omega$. Without restriction assume that $\omega(h)=\omega(k)=0$
and $\dot\gamma_0=\dot\eta_0$. Then (\ref {relent:initial2}) implies that
\be
(T_{_\Omega} (h-k)\Omega,T_{_\Omega}  x^*\Omega)&=&0,
\qquad x\in\Malg.
\nnee
Take $x=h-k$. Then it follows that $T_{_\Omega} (h-k)\Omega=0$.

\endproof

The above lemma shows that the map 
\be
\dot\gamma_0\mapsto T_{_\Omega} (h-\omega(h))\Omega
\label{metric:identif}
\ee
is one-to-one and identifies  the tangent vector $\dot\gamma_0$ with the vector  $T_{_\Omega} h\Omega$
in the Hilbert space $\mathscr H$.

Expression (\ref {metric:temp}) defines a bilinear form. This follows from

\begin{lemma}
The map (\ref {metric:identif}) is linear. 
\end{lemma}

\beginproof
Let $\gamma$ be an exponential arc with generator $h$ in $\Malg$.
Then $t\mapsto\gamma_{\epsilon t}$ is an exponential arc with generator $\epsilon h$
for any $\epsilon$ in $[-1,1]$ and the tangent vector is $\epsilon\dot\gamma_0$.
Hence, (\ref {metric:identif}) maps $\epsilon\dot\gamma_0$ onto $\epsilon T_{_\Omega} h\Omega$.

Next consider a pair of exponential arcs $\gamma$ and $\eta$ with generators
$k$, respectively $h$, in $\Malg$ and with $\gamma_0=\eta_0=\omega$.
Let $\theta$ denote the exponential arc with generator $h+k$. It exists by Theorem \ref {relent:theorem}.
The state $\theta_t$ can then be written as
\be
\theta_t(x)&=&(x\Phi_{th+tk},\Phi_{th+tk})
\nnee
with $\Phi_{th+tk}$ the unique element in the natural positive cone representing the state $\theta_t$.
Use now (\ref {relent:expans}) to write
\be
\theta_t(x)&=&\omega(x)+\frac t2\int_0^1\upd u\,(x\Delta_{_\Omega}^{u/2}(h+k)\Omega,\Omega)\cr
& &\qquad+
\frac t2\int_0^1\upd u\,(x\Omega,\Delta_{_\Omega}^{u/2}(h+k)\Omega)+\mbox{ o}(t).
\nnee
This implies
\be
\dot\theta_0&=&\dot\gamma_0+\dot\eta_0.
\nnee

Both observations together prove linearity of the map (\ref {metric:identif}).

\endproof

\begin{proposition}
\label{metric:prop:nondeg}
Expression (\ref {metric:temp}) defines a non-degenerate scalar product on the 
space of tangent vectors of the form $\dot\gamma_0$ with $\gamma$ an exponential arc as
constructed in Theorem \ref {relent:theorem}.
\end{proposition}

\beginproof
The two lemmas show that (\ref {metric:temp}) is a well-defined bilinear form.
Positivity of the form is clear. The symmetry follows from (\ref {metric:temp}).
It remains to be shown that it is non-degenerate.

Assume that $(\dot\gamma_0,\dot\gamma_0)=0$. This implies
\be
T_{_\Omega} (h-\omega(h))\Omega&=&0,
\nnee
with $h$ the generator of $\gamma$. 
The operator $T_{_\Omega} $ is invertible --- see the proof of  Lemma II.2 of \inlinecite{NVW75}.
Hence, it follows that
\be
(h-\omega(h))\Omega&=&0.
\nnee
Because $\Omega$ is separating for $\Malg$ it follows that $h$ is a multiple of the identity.
The latter implies that $\dot\gamma=0$.
\endproof

\section{Dual geometries}

\hskip 2cm\parbox{\pbsize}
{\em
The geodesics of the e-connection are the exponential arcs.
In the m-connection the geodesics are made up by convex combinations of a pair of states.
The m- and e-connections are each others dual w.r.t.~the metric of Section \ref {sect:metric}.
}

Consider two states $\omega$ and $\phi$ in the manifold $\Mo$. 
The tangent vector
\be
\dot\gamma_t=\frac{\upd\,}{\upd t}\gamma_t=\phi-\omega,
\qquad 0<t<1,
\nnee
is independent of $t$.  Hence, it is a geodesic for the connection in which
all parallel transport operators are taken equal to the identity operator.
Note that the tangent space $T_\omega\Mo$ coincides with the space of $\sigma$-weakly continuous
linear functionals $\chi$ satisfying $\chi(1)=0$ and hence is everywhere the same.
This connection is by definition the m-connection.

For $t$ in $(0,1)$ the tangent vector $\dot\gamma_t$ belongs to the subspace ${\mathscr T}_{\gamma_t}$
of the tangent space $T_\omega\Mo$ which is introduced in Section \ref {sect:approx}.
Conversely, every vector $\chi$ in ${\mathscr T}_{\gamma_t}$ is the tangent vector
of an m-geodesic passing through the point $\gamma_t$.
However, this does not imply that through parallel transport $\Pi(\gamma_t\mapsto \gamma_s)$
the space ${\mathscr T}_{\gamma_t}$ maps onto the space ${\mathscr T}_{\gamma_s}$.

The transport operators $\Pi^*$ of the dual geometry are defined by
\be
\left(\Pi(\phi\mapsto\omega)V,\Pi^*(\phi\mapsto\omega)W\right)_\omega
&=&
\left(V,W\right)_\phi.
\nnee
In this expression $V$ and $W$ are vector fields and $(\cdot,\cdot)_\omega$ is the scalar product 
defined in the previous section and evaluated at the point $\omega$ of the manifold $\Mo$.

It is shown below that any exponential arc $\gamma$ is a geodesic for this dual geometry.
To do so we have to show that 
\be
\Pi^*(\gamma_s\mapsto\gamma_t)\dot\gamma_s&=&\dot\gamma_t.
\nnee
The tangent vector $\dot\gamma_t$ at $t=0$ is given by (\ref {relent:initial2}). Its value for arbitrary $t$
is given by the following proposition.

\begin{proposition}
Let $\gamma$ denote an exponential arc $\gamma$ with generator $h$ belonging to $\Malg$.
Let $\Phi_t$ be the normalized vector in the natural positive cone $\mathscr P$ representing the state $\gamma_t$.
The derivative $\dot\gamma_t$ is given by
\be
\dot\gamma_t(x)&=&\left(T_{\Phi_t} h{\Phi_t},T_{\Phi_t} [x-\gamma_t(x)]^*\right){\Phi_t}),
\qquad x\in\Malg.
\label{dual:deriv}
\ee
\end{proposition}

\beginproof

The state ${\gamma_1}$ is connected to $\omega$ by the exponential arc with generator $h$
and $\gamma_t$ is connected to $\omega$ by the exponential arc with generator $th$.
Let 
\be
\psi_s&=&\gamma_{(1-s)t+s}.
\nnee
It follows from Proposition \ref {arcs:prop:general} that $s\mapsto \psi_s$
is an exponential arc with generator $(1-t)h$  connecting $\gamma_t$ to $\gamma_1$.
Application of (\ref {relent:initial2}) to the latter arc gives
\be
\dot\psi_0(x)=
\frac{\upd\,}{\upd s}\bigg|_{s=0}\psi_s(x)&=&(1-t)\,(T_\Psi h\Psi,T_\psi (x-\omega(x))^*\Psi)
\label{dual:temp}
\ee
with $\Psi=\Phi_t$.
This implies (\ref {dual:deriv}) because $\dot \psi_0=(1-t)\dot\gamma_t$.

\endproof

\begin{theorem}
Any exponential arc $\gamma$ with generator $h$ in $\Malg$ is a geodesic for the dual
of the m-connection w.r.t.~the metric introduced in Section \ref {sect:metric}.
\end{theorem}

\beginproof

Let $t\mapsto\phi_t$ be an exponential arc with generator $k$ in $\Malg$
such that $\phi_0=\gamma_t$.
Fix $t$ in $[0,1]$ and let $\Phi_t$ denote the normalized element of the natural positive cone $\mathscr P$
representing the state $\gamma_t$.
Let $\eta$ be an exponential arc with generator $k$ starting at $\gamma_t$,
i.e.~$\eta_0=\gamma_t$.
Because $\Pi(\gamma_s\mapsto\gamma_t)$ is the identity the definition of the dual transport operator yields
\be
\left(\dot\eta_0,\Pi^*(\gamma_s\mapsto\gamma_t)\dot\gamma_s\right)_{\gamma_t}
&=&
\left(\dot\eta_0,\dot \gamma_s\right)_{\gamma_s}\cr
&=&
\left(T_{\Phi_t}(k-\gamma_t(k))\Phi_t,T_{\Phi_t}(l-\gamma_t(l))\Phi_t\right),
\nnee
with $l$ the generator of the arc $s\mapsto \gamma_{(1-s)t+s}$. It equals $l=(1-t)h$.
This last expression equals
\be
&=&
\left(\dot\eta_0,\dot\gamma_t\right)_{\gamma_t}.
\nnee
By proposition \ref {metric:prop:nondeg} the scalar product $(\cdot,\cdot)_{\gamma_t}$
is non-degenerate. Therefore one can conclude that
\be
\Pi^*(\gamma_s\mapsto\gamma_t)\dot\gamma_s&=&\dot\gamma_t.
\nnee
This shows that the exponential arc $\gamma$ is a geodesic for the dual of the m-connection.

\endproof

\section{Finite-dimensional submanifolds}
\label{sect:findim}

\hskip 2cm\parbox{\pbsize}
{\em
A finite set of linearly independent generators is shown to define a finite-dimensional submanifold
in which all states are connected to the reference state by an exponential arc.
The submanifold defined in this way is a dually flat quantum statistical manifold.

  }

  Let $\omega$ be the reference state of $\Mo$. It is a vector state with cyclic and separating vector $\Omega$.
Choose an independent set of selfadjoint operators $h_1,\cdots,h_n$ in $\Malg$. 
By Theorem \ref {relent:theorem} there exists an exponential arc $\gamma$
with generator $h=\theta^ih_i$ connecting some state $\gamma_1$ in $\Mo$ to
the state $\gamma_0=\omega$.
A parameterized family of states $\omega_\theta$, $\theta\in\Ro^n$
is now defined by putting $\omega_\theta=\gamma_1$.
These states form a submanifold of $\Mo$.

From the definition of an exponential arc one obtains immediately
that for any $\psi$ in $\Mo$
\be
D(\psi||\omega_\theta)
&=&
D(\psi||\omega)+D(\omega||\omega_\theta)+\theta^i\left(\omega(h_i)-\psi(h_i)\right).
\label{geom:theo:a}
\ee
Take $\psi=\omega_\theta$ in this expression to find
\be
\theta^i\omega(h_i)
\,\le\,
D(\omega||\omega_\theta)+\theta^i\omega(h_i)
&=&
\theta^i\omega_\theta(h_i)-D(\omega_\theta||\omega)
\,\le\,
\theta^i\omega_\theta(h_i).
\label{geom:Phidef}
\ee
Hence, the quantity $\theta^i\omega(h_i)$ is maximal if and only if $\omega_\theta$ equals 
the reference state $\omega$.

\begin{proposition}
Dual coordinates $\eta_i$ are defined by
\be
\eta_i&=&\omega_\theta(h_i).
\nnee
They satisfy
\be
\frac{\partial\eta_i}{\partial\theta^j}&=&(\partial_i,\partial_j)_\theta
\nnee
with $(\cdot,\cdot)_\theta$ equal to the scalar product  $(\cdot,\cdot)_{\omega_\theta}$
introduced in Section \ref {sect:metric} and with basis vectors $\partial_i$
equal to $\partial\omega_\theta/\partial\theta^i$.
\end{proposition}

\beginproof
Introduce the path $\gamma^{(i)}$ defined by
\be
\gamma^{(i)}:\,t\mapsto \omega_{\theta+tg_i}.
\nnee
It satisfies 
\be
\frac{\partial\,}{\partial\theta^i}\omega_\theta=\partial_i=\dot\gamma^{(i)}(0).
\nnee
By definition is $\omega_{\theta+g_i}$ the end point of the exponential arc with generator 
$(\theta^j+g^j_i)h_i$. From Proposition \ref {relent:prop:additive} it then follows that
$\gamma^{(i)}$ is an exponential arc with generator $h_i$
connecting $\omega_{\theta+g_i}$ to $\omega_{\theta}$. 
These arcs $\gamma^{(i)}$ are used in the calculation that follows.

The definition of the scalar product at the beginning of Section \ref{sect:metric} gives
\be
(\partial_i,\partial_j)_{\omega_\theta}
&=&
(\dot\gamma^{(i)}(0),\dot \gamma^{(j)}(0))_{\omega_\theta}\cr
&=&-\frac{\partial\,}{\partial s}\frac{\partial\,}{\partial t}\bigg|_{s=t=0}D(\gamma^{(i)}_s||\gamma^{(j)}_t)\cr
&=&-\frac{\partial\,}{\partial s}\bigg|_{s=0}\frac{\partial\,}{\partial\theta^j}D(\gamma^{(i)}_s||\omega_\theta)\cr
&=&
\frac{\partial\,}{\partial s}\bigg|_{s=0}\gamma^{(i)}_s(h_j)\cr
&=&
\partial_i(h_j)\cr
&=&
\frac{\partial\eta_j}{\partial\theta^i}.
\nnee

\endproof

\begin{corollary}
There exists a potential $\Phi(\theta)$ such that
\be
\eta_i&=&\frac{\partial\Phi}{\partial\theta^i}.
\label{geo:Phiexist}
\ee
\end{corollary}

This follows because the scalar product is symmetric so that
\be
\frac{\partial\eta_j}{\partial\theta^i}=(\partial_i,\partial_j)_\theta=(\partial_j,\partial_i)_\theta
=\frac{\partial\eta_i}{\partial\theta^j}.
\nnee
This symmetry is a sufficient condition for the potential $\Phi(\theta)$ to exist.

Consider the following generalization of the potential introduced in Section \ref {sect:scalar}.
\be
\Phi(\theta)&=&D(\omega||\omega_\theta)+\theta^i\omega(h_i).
\label{geo:Phidef}
\ee
Apply (\ref {relent:deriv}) to the exponential arc $\gamma^{(i)}$ which connects
$\omega_{\theta+g_i}$ to $\omega_\theta$ to find
\be
\frac{\partial\,}{\partial\theta^i} D(\omega||\omega_\theta)
&=&
\omega_\theta(h_i)-\omega(h_i).
\nnee
This implies that $\Phi(\theta)$ satisfies (\ref {geo:Phiexist}).

One concludes that the selection of an independent set of selfadjoint operators $h_1,\cdots,h_n$ in $\Malg$
defines a parameterized statistical model $\theta\mapsto\omega_\theta$ of states on the von Neumann algebra $\Malg$.
An obvious basis in the tangent plane $T_{\omega_\theta}\Mo$ is formed by the derivative operators $\partial_i$.
The scalar product $(\partial_i,\partial_j)_{\omega_\theta}$ introduced in Section \ref {sect:metric}
starting from the relative entropy of Araki defines a Hessian metric on the tangent planes.
Exponential arcs are geodesics for the e-connection which is the dual of the m-connection. 

\section{Discussion}

\begin{itemize}
 \item 

The manifold $\Mo$ under consideration consists of vector states 
on a sigma-finite von Neumann algebra $\Malg$ in its standard representation. 
Such a manifold has nice properties described
by Tomita-Takesaki theory and hence is an obvious study object when exploring 
quantum statistical manifolds in an infinite-dimensional setting.
Particular attention is given in the present work on the definition of the tangent planes.
This is also a point of concern in the commutative context of manifolds of probability measures.
See for instance the approach of \inlinecite{PS95}. A convenient choice for the tangent space
$T_\omega\Mo$ at the state $\omega$ in the manifold $\Mo$ 
is to take it equal to the space
of all $\sigma$-weakly continuous Hermitian linear functionals $\chi$ on $\Malg$ vanishing on the identity operator $\Io$.
However, it is well possible that the equivalence class of smooth curves through $\omega$
with initial tangent equal to a given $\chi$ is empty.
Approximate tangent planes are considered as an alternative in Section \ref {sect:approx}. 
They form a subspace of $T_\omega\Mo$ as defined above.
Never the less, the initial tangent vectors of Fr\'echet-differentiable paths
starting at $\omega$ belong to the approximate tangent space.
It is not clear whether the initial tangents of exponential arcs 
 are dense in the approximate tangent space w.r.t.~the inner product of Section \ref {sect:metric}. 
Further research is needed at this point.

\item
A new definition of exponential arcs is given.
It depends on the choice of a divergence function/relative entropy
defined on pairs of points in the manifold
and on the choice of a generator which is a linear functional defined on a domain in the manifold.
It is general enough to cover different approaches that one can follow to
solve the non-uniqueness problem of
the Radon-Nikodym derivative in the context of non-commutative probability.
Never the less one can prove in full generality nice properties such as uniqueness of the generator,
existence of a scalar potential, Pythagorean relations.
Additivity of generators when composing exponential arcs is shown in the specific context
of Araki's relative entropy. See Proposition \ref {relent:prop:additive}.

\item
The second half of the paper focuses on the relative entropy of Araki.
Only exponential arcs with bounded generators belonging to the von Neumann algebra are considered.
This suffices to reach the goal of replacing the existing approach based on density matrices
and Umegaki's relative entropy. However, the solution of the problem mentioned above regarding
the extent of the tangent spaces most likely requires the handling of unbounded generators.

\item
The scalar product of Bogoliubov presented in Section \ref{sect:metric} is used extensively in
Linear Response Theory, also known as Kubo-Mori theory.
Its link with the KMS condition of Section \ref {sect:kms} is not highlighted in the present text.
It is tradition in Kubo-Mori theory and more generally in
Statistical Mechanics to focus on a small number of variables.
It is shown in Section \ref {sect:findim} that the selection of a finite number of variables
defines a quantum statistical manifold supporting Amari's dually flat geometry.

\end{itemize}

\vskip 2cm

\bibliographystyle{plain} 
\small
\bibliography{manuscript}

\end{document}